\documentclass[10pt,a4paper,twoside]{article}
\usepackage{JISE}
\usepackage{graphicx}
\usepackage{times}
\usepackage{mathptmx}

\title{Citizens' Emotion on GST: A Spatio-Temporal Analysis over Twitter Data}
\author{Deepak Uniyal, Ankit Rai\\
  {\small\em Department of Computer Science and Engineering} \\
  {\small\em Graphic Era Deemed To Be University }\\
  {\small\em Dehradun, Uttarakhand}\\
  {\small\em E-mail: \{deepak.uniyal08; ankitrai326;\}@gmail.com}
}

\authorrunning{Deepak Uniyal, Ankit Rai}
\titlerunning{Citizens' Emotion on GST: A Spatio-Temporal Analysis over Twitter Data}
\usepackage{graphicx}
\usepackage{hhline}
\usepackage{subfig}

\begin{document}
\maketitle

\begin{abstract}
People might not be close-at-hand but they still are - by virtue of the social network. The social network has transformed lives in many ways. People can express their views, opinions and life experiences on various platforms be it Twitter, Facebook or any other medium there is. Such events constitute of reviewing a product or service, conveying views on political banters, predicting share prices or giving feedback on the government policies like Demonetization or GST.
These social platforms can be used to investigate the insights of the emotional curve that the general public is generating. This kind of analysis can help make a product better, predict the future prospects and also to implement the public policies in a better way. Such kind of research on sentiment analysis is increasing rapidly.
In this research paper, we have performed temporal analysis and spatial analysis on 1,42,508 and 58,613 tweets respectively and these tweets were posted during the post-GST implementation period from July 04, 2017 to July 25, 2017. The tweets were collected using the Twitter streaming API. A well-known lexicon, National Research Council Canada (NRC) emotion Lexicon is used for opinion mining that exhibits a blend of eight basic emotions i.e. joy, trust, anticipation, surprise, fear sadness, anger, disgust and two sentiments i.e. positive and negative for 6,554 words.
\end{abstract}

\section{INTRODUCTION}

Goods and Services Tax (GST) is a concept of One-Nation, One-Tax, a Single Tax policy which is applied to all the goods and services at a national level. Introduction of GST brings all the taxes under the umbrella of of a uniform Tax Base. GST was rolled out by the Union government on the recommendations of the GST Council, which is the key decision-making body of GST regime.
The Council is chaired by the Union Finance minister, Arun Jaitley. As per Article 279A of the amended Constitution, the GST Council is a joint forum of the Center and the states \footnote{http://www.gstcouncil.gov.in/}. 

The first draft of the GST bill was brought up in the parliament in the year 1999 under the leadership of the then PM of India Mr. Atal Bihari Vajpayee. However, it took almost two decades to  implement it. The GST Bill was implemented at midnight (June 30th, 2017 - July 1st, 2017) by the President of India and the Government of India \footnote{https://en.wikipedia.org/wiki/Goods\_and\_Services\_Tax\_(India)}.

Before the GST regime, various indirect taxes were levied by the Central and the State governments. Indian consumers had to pay taxes like Value Added Tax (VAT) or Local Sales Tax, Service Tax, Central Excise Duty, Customs Duty etc. Now, the GST will subsume almost all the indirect taxes of the Center and the State governments. The new tax system is fairly simple vis-a-vis its predecessor. The tax slabs  \cite{C7} for GST in India are 0\%, 5\%, 12\%, 18\%, and 28\%, depending on the type of product/service. 
For example, Zero (0\%) is for all essential items including food grains and Five (5\%) for items of mass consumption including essential commodities. There are few products like Petroleum, alcoholic drinks, electricity, and real estate that are taxed separately by the respective state governments, as per the previous tax regime. 
There are three forms of GST implemented in India - Central-GST (CGST), State-GST (SGST), and Interstate-GST (IGST). A seller has to collect both CGST and SGST from the buyer in Intra-State transactions, whereas he has to collect IGST in an Inter-State transaction.

Prior to India, more than 160 countries have already implemented the GST form of Tax Structure. The GST in India, in its current form, is a bit distinct and complex than the rest of countries. Unlike other nations, GST in India will be charged at different rates depending upon the category, value or state of the business. In Asia Pacific region, more than 40 kinds of GST models are operational that make a diverse set of rules and regulations. 

It was considered as one of the biggest tax reforms of Indian history, which was expected to boost overall growth of the economy. Suitably, the repercussions were huge too.
Introduction of GST created a lot of ripples on major social media platforms including Facebook and Twitter. People engaged in discourse, expressing all kinds of opinions on taxation system in India and the plausibility of a brand new idea. 

Researchers use this data to understand the insight of public sentiments, emotions or opinions on an event and present a report to the concerned authority in case any further action is required e.g. to make a better product, predict the target price of the shares or use public inputs to implement the government policy in a better way. Such analysis where the data collected from social media platforms is mined to analyse the public sentiments is called sentiment analysis or opinion mining.

The first step in Sentiment Analysis is to collect data from social media accounts, and preprocess it to extract meaningful information from the data.
Opinion mining can be performed by understanding the sentiments of different words in clean and preprocessed data. To deduce the emotions and sentiments we have used NRC - Lexicon which consists of 2554 words, NRC - Lexicon lists down eight emotions (joy, trust, anticipation, surprise, fear sadness, anger and disgust) and two sentiments (positive and negative). 

We performed two type of analysis - Temporal and Spatial. In Temporal analysis, we calculated the frequency of tweets per hour and presented the varying emotions graphically while in Spatial analysis, we found the location of tweets and tried to show the degree of acceptance, region wise.
Data were collected from July 4th to July 25th, 2017. The purpose of collecting data over a month is to make a fair judgement about the sentiments of people over a long period of time.

\section{RELATED WORK}
Sentiment analysis can be defined as the process of automatically determining the sentiments, views, emotions, opinions and attitudes towards a social event.
According to previous \cite{C1} opinions, views, sentiments are used interchangeably but still there is some difference between them. 

According to \cite{C2} sentiment analysis is about finding out opinions, identifying the sentiments people express and then classifying their polarity which could be positive, negative or neutral.
There are two types of approaches for sentiment analysis -  machine learning techniques (supervised and unsupervised) and lexicon based approach. Lexical based approach is further divided in to Dictionary based approach and Corpus based approach. Dictionary based approach relies on the terms collected and annotated manually and grows by adding synonyms and antonyms of the dictionary words, whereas corpus based approach uses a domain specific dictionary.

We have studied some of the past research that have been performed in the sentiment analysis in various fields \cite{C3} \cite{C4}. Some of the researchers are using standard datasets that are already annotated whereas some are using their own datasets for analysis.

In \cite{C5}, sentiment analysis has been performed on  104,003 tweets which were collected before the German national election between August 13th and September 19th, 2009. The analyses show that a small section of people are dominating the social media and more than 40\% of the messages or posts are made by only 4\% of the people. The analyses also indicate a close correspondence between political discussions on public platforms and offline scenario. 

In\cite{C6} a survey is presented that talks about different features in news content like linguistic features which are helpful in detecting fake news. There are many other similar tasks, like spam detection, rumour classification, on truth discovery that paper has discussed.

A similar paper is \cite{C7} which has presented a work that identifies fake tweets by using various features from tweets and twitter accounts.

In \cite{C8} a system has been developed for real-time analysis of public sentiments towards US presidential election 2012. The system analyses and visualizes over 36 million tweets that shows number of real time tweets every minute and tweets with different polarity of sentiments every five minutes.

Reserach \cite{C9} uses Naive Bayes Algorithm to analyse public sentiments on approximately 20000 tweets collected from 27th June 2017 to 07th July 2017. The analysis shows that approximately 52\% tweets are positive about the new taxation system in India while the rest of the tweets, negative and neutral are also quite significant in number.

In \cite{C10} \cite{C11} a technique is used to process and analyse a huge amount of real time data using Hadoop cluster and Naive Bayes approach. The paper focuses on the speed of the sentiment analysis rather than accuracy of the analysis. It processes the data by removing stop words, converting unstructured data to structured data and replacing emotions to their corresponding words.

In \cite{C12} sentiment analysis has been performed on over 2,50,000 tweets mentioning  \#Microsoft, \#Windows, \$MSFT etc. using supervised machine learning approach to find out the correlation between stock market movements and sentiments in the tweets. 

In\cite{C13} author finds correlation between Cricketers' performances and fans' emotions. Their work shows that fans' emotions depend on players' performance in the tournament. In\cite{C14} the author performs the Geo-spatial sentiment analysis for the UK-EU referendum over the Twitter data. It analyses the data to find out the most talked about British politicians and the public sentiments for them. We have used the same approach to remove the noise from tweets as discussed in \cite{C15} research.

In \cite{C16} authors have witnessed correlations between the good performance with positive emotions and bad performance with negative emotions. Besides this, they found Trust (a type of positive emotion) as a most entangled emotion corresponding to each performance. Moreover, trading price of commercial brands are found to have transitive relationship with their brand ambassadors' performance in the match. As a reference, they have used Google Trends to verify the influence of players performance all over the globe. 

\section{Methodology}
\subsection{Data Collection}
Before we perform the sentiment analysis, we need to collect the data. The most important thing before we collect data is to decide the source of data and the time period for which we need to collect it. We selected the Twitter as a data source, as nowadays Twitter is very popular platform worldwide for public to express their views, opinions and sentiments like happiness, celebration, anger, surprise or fear etc. towards an event.

Dataset can be collected in three time periods - before the event has occurred, during the event and after the event has occurred.
We chose to collect and analyse the data three days after the GST was implemented as by that time people would have got some information about the GST via social media, TV debates and other sources which decreases the probability of being biased or prejudices towards the policy.

We collected 1,63,372 tweets over 22 days with the help of twitter streaming API by using some popular and trending hashtags, which we were able to identify after manual scan of popular hashtags. After data preprocessing and cleaning process we removed more than 20000 tweets as they lacked the meaningful information. Finally 1,42,508 tweets were further considered for sentiment analysis. {TABLE-1} shows the number of tweets collected during 22 days starting from 4th July, 2017 to 25th July, 2017.
\begin{table}[tb]
	\caption{Number of Tweets Per Day (July 2017)}
	\begin{center}
		\begin{tabular}{|c|c|c|c|}
			\hline
			\cline{2-4} 
			\textbf{Date} & \textbf{\textit{No of Tweets}}& \textbf{\textit{Date}}& \textbf{\textit{No of Tweets}} \\
			\hline
			4\textsuperscript{th} July& \textit{13025}& 5\textsuperscript{th} July& \textit{16171} \\
			\hline
			6\textsuperscript{th} July& \textit{7644}& 7\textsuperscript{th} July& \textit{10778} \\
			\hline
			8\textsuperscript{th} July & \textit{15834}& 9\textsuperscript{th} July& \textit{12964} \\
			\hline
			10\textsuperscript{th} July & \textit{6170}& 11\textsuperscript{th} July& \textit{5240} \\
			\hline
			12\textsuperscript{th} July & \textit{1772}& 13\textsuperscript{th} July& \textit{7821} \\
			\hline
			14\textsuperscript{th} July & \textit{4241}& 15\textsuperscript{th} July& \textit{5869} \\
			\hline
			16\textsuperscript{th} July & \textit{4099}& 17\textsuperscript{th} July& \textit{6801} \\
			\hline
			18\textsuperscript{th} July & \textit{4782}& 19\textsuperscript{th} July& \textit{4439} \\
			\hline
			20\textsuperscript{th} July & \textit{1892}& 21\textsuperscript{th} July& \textit{2034} \\
			\hline
			22\textsuperscript{nd} July & \textit{2252}& 23\textsuperscript{rd} July& \textit{3677} \\
			\hline
			24\textsuperscript{th} July & \textit{4419}& 25\textsuperscript{th} July& \textit{644} \\
			\hline
			\hline
			\multicolumn{2}{c}{Total Tweets} & \multicolumn{2}{c}{\textit{1,42,508}} \\
			\hline
			\hline
		\end{tabular}
		\label{tab1}
	\end{center}
\end{table}

\subsection{Data Preprocessing and Cleaning}
Data Preprocessing is a data mining technique that transforms raw data into a meaningful information which is understandable and ready for mining. This phase is very critical and a crucial step in data mining as it hugely impacts the analysis. There are high chances that unprocessed data remains inconsistent and noisy which might result into incorrect analysis. Number of previous study has fallow different approaches to tackle with noisy content[8][10][11][12][13]. To make the data uniform and clean we pre-process it by removing all the characters, words and phrases that are less significant or carry less weightage in sentiment analysis.

We passed each and every tweet through a python program that performs the preprocessing as shown in Fig-1 through a sample tweet. It starts with the removal of all the hashtags, mentions and URLs which is further processed by removing all the punctuations, single \& double characters and recurring characters. Finally it removes all the stop words from all the tweets against the list, commonly used as stop words list to keep only meaningful data. In the final step of data preprocessing we have trimmed the tweets and then removed all the blank tweets from the data corpus. As a result, we have the dataset containing tweets, which will be more significant in the analysis and will carry some sentiments or emotions. 

\begin{figure*}[!t]
	\centering
	\subfloat[Preprocessing and Sentiment Analysis -1]
	{\includegraphics[width=6.3 cm,height=4cm]{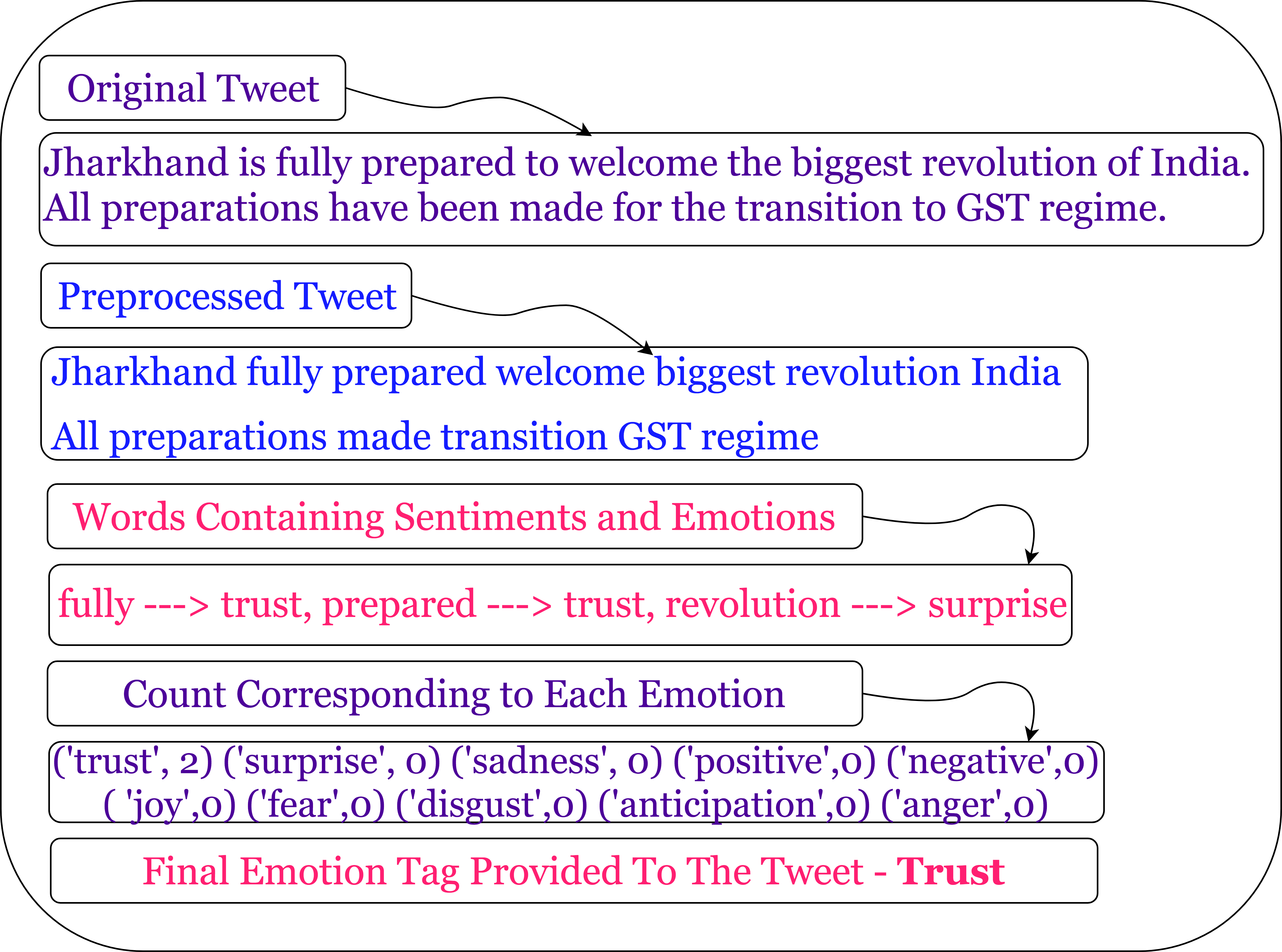}%
		\label{f1}}
	\subfloat[Preprocessing and Sentiment Analysis -2]
	{\includegraphics[width=6.3 cm,height=4cm]{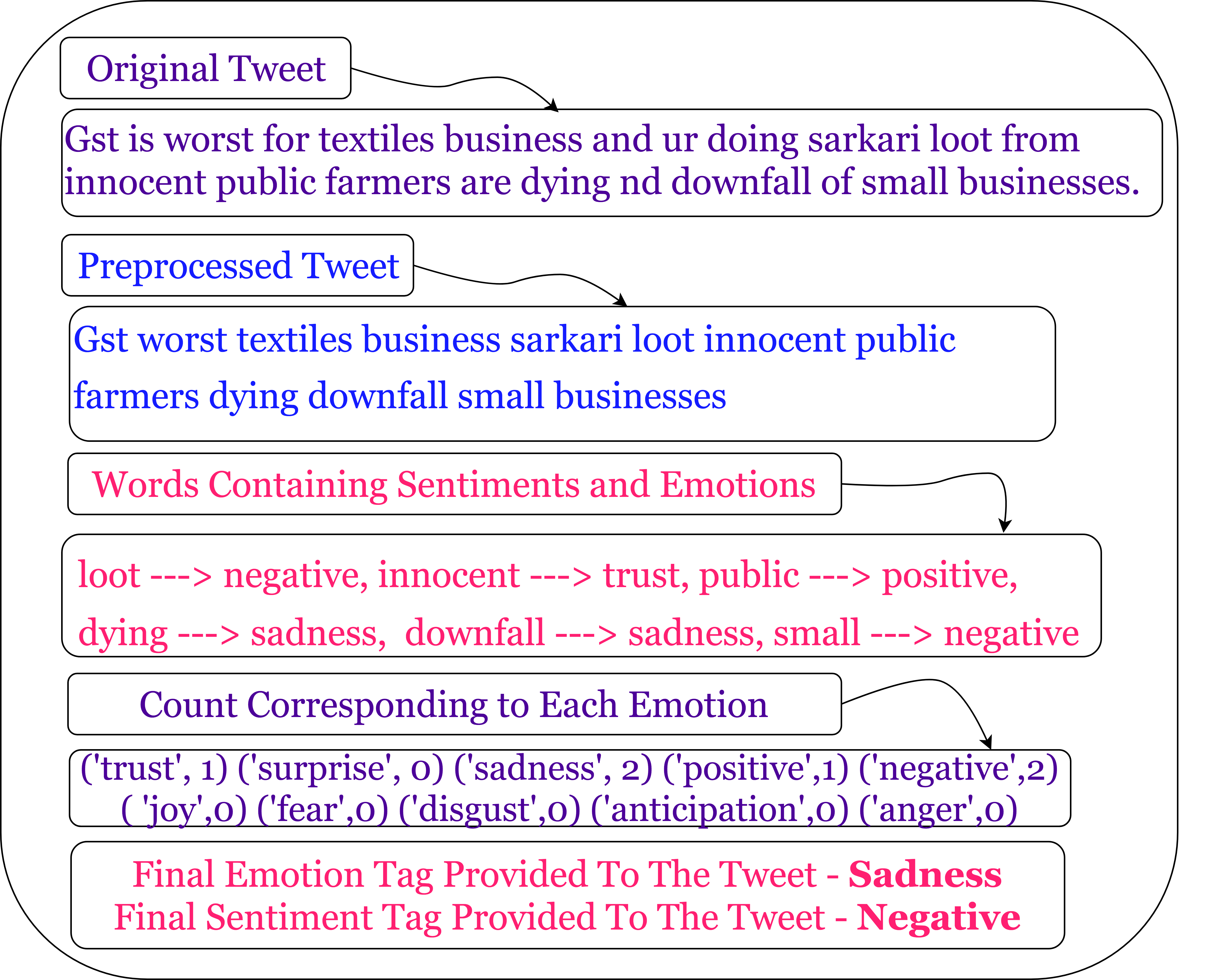}%
		\label{f2}}
	\caption{Showing the Data Preprocessing and Sentiment Analysis Process on             	Sample Tweets }
\end{figure*}

\subsection{Sentiment Analysis}
After the data has been preprocessed, it is ready for analysis with the tweets containing all the words with more weightage or sentiments. In this paper, we have performed analysis on the basis of time and location on 1,42,508 and 58,613 tweets respectively.

\begin{figure}[!t]
	\centering
	\subfloat[Sentiments ]
	{\includegraphics[width=6.3 cm,height=4cm]{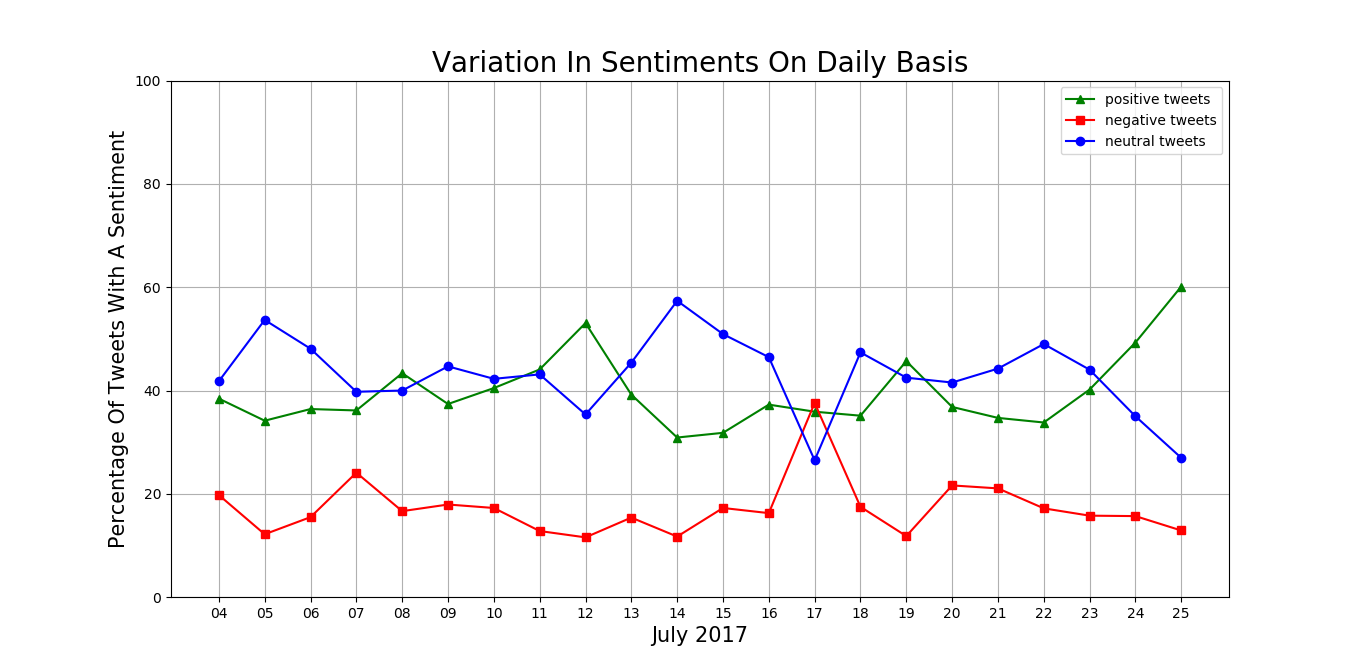}%
		\label{f1}}
	\subfloat[Emotions]
	{\includegraphics[width=6.3 cm,height=4cm]{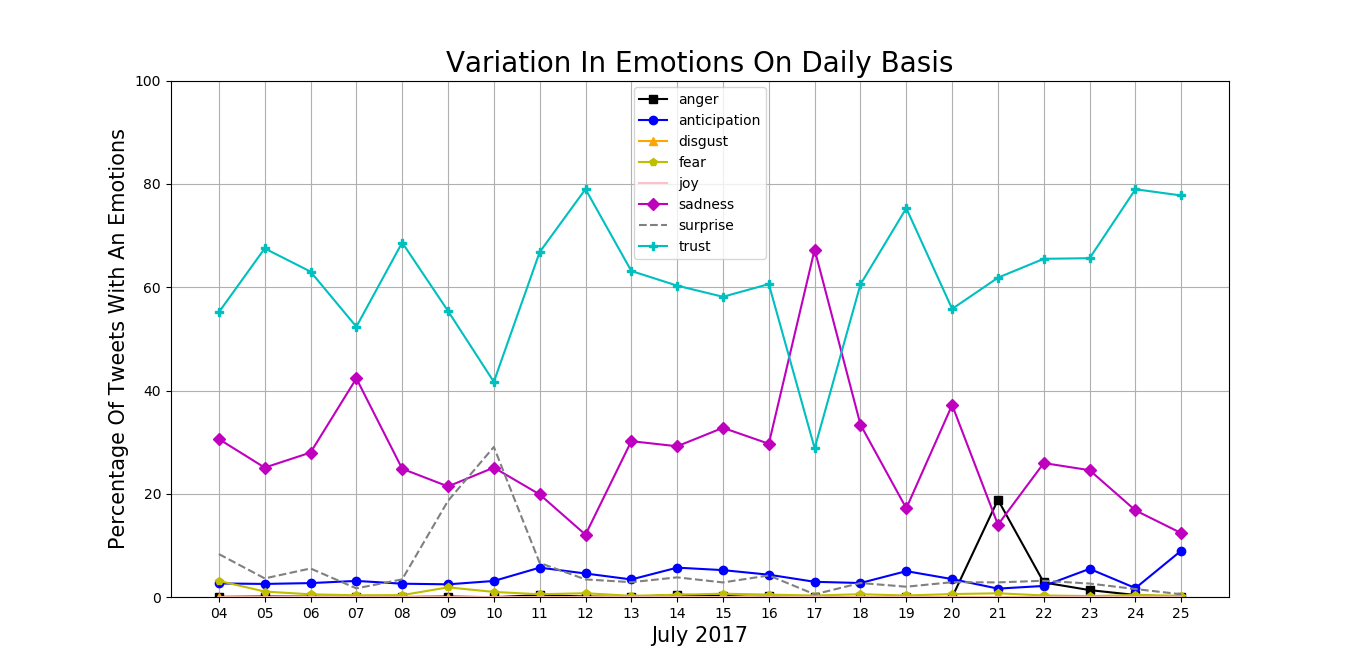}%
		\label{f2}}
	\\
	\subfloat[Hourly Frequency]
	{\includegraphics[width=10 cm,height=4.5cm]{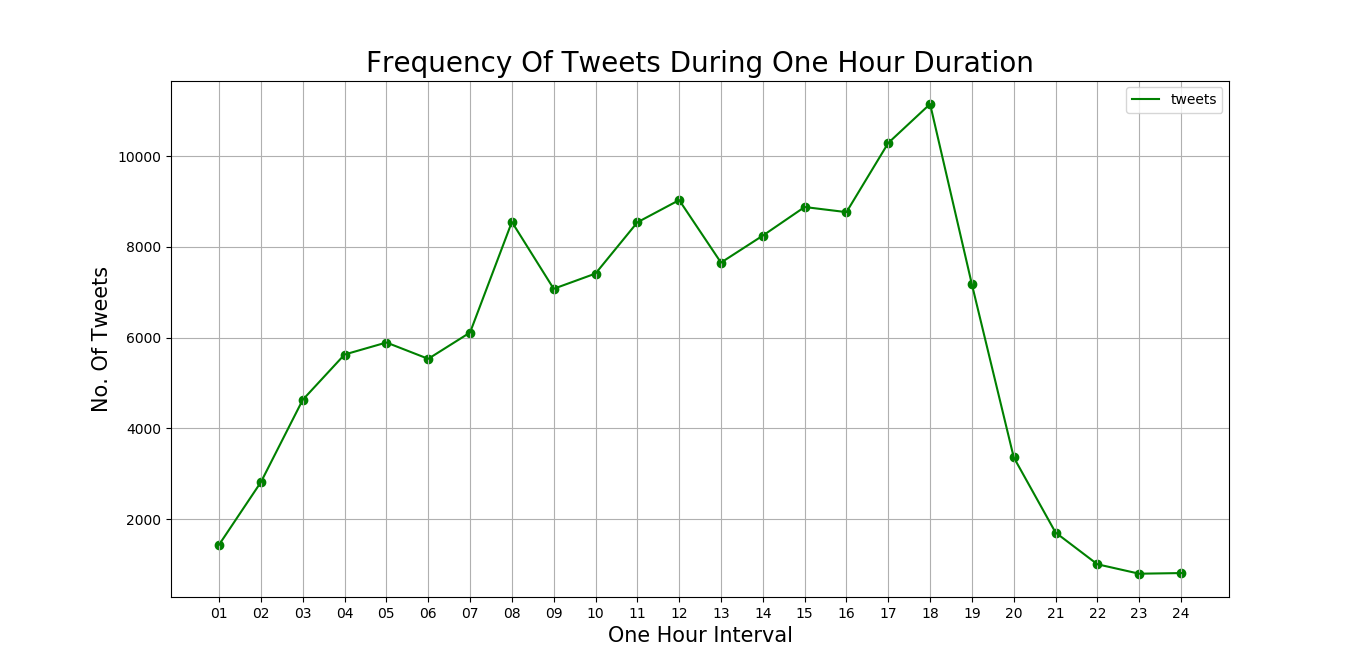}%
		\label{f2}}
	\caption{Showing the varying Sentiments, Emotions and Hourly Frequency of Tweets over 22 days in July 2017}
\end{figure}

\subsubsection{Temporal Analysis}
In case of temporal or time based analysis, the sentiment analysis has been performed on the data on an hourly and a daily basis. On preprocessed tweets, we have calculated the number of tweets with corresponding emotions (joy, trust, anticipation, surprise, fear sadness, anger and disgust) and sentiments (positive and negative), as shown in Fig-1 by a sample tweet, using NRC emotion lexicon.

To calculate the number of tweets and corresponding emotion or sentiment, we have applied the following methodology -
\begin{itemize}
	\item Tag each word of the tweet with the corresponding emotion and sentiment as per NRC emotion lexicon. If a word corresponds to more than one emotion, tag the word with each one of them.
	\item Tag a tweet as neutral, if none of the word in a tweet has been tagged to any of the emotions or sentiments.
	\item Mapping emotions, sentiments and neutral responses to their counts in the tweet.
	\item Provide an emotion, a sentiment or a neutral tag to the tweet on the basis of maximum count or score we achieved in the previous step.
\end{itemize}
After applying the above methodology, all the tweets are tagged to an emotion, a sentiment or as neutral tweets. Finally we have created a mapping for emotions, sentiments, and neutral tweets to their counts per day starting from July 04, 2017 to July 25, 2017.

The graph in Fig-2(a) shows that most of the tweets are either neutral or positive and negative tweets are approximately half of the positives, which indicates that most of the people are positive about the modified taxation system of India. 

We have also shown a graph for emotions and corresponding counts on a daily basis in Fig-2(b). This graph represents that most of the tweets indicate trust, followed by sadness, anticipation, surprise and fear in that order.

In both the graphs we can see a spike on 17\textsuperscript{th} July which indicates a sudden jump in negativity and sadness in public emotions, but that again rebounds to higher positives and trustworthy tweets from the next day.
\begin{table}[tb]
	\caption{Number of Mentions Per User (July 2017)}
	\begin{center}
		\begin{tabular}{|l|l|r|}
			\hline
			\cline{1-3} 
			\textbf{User}&\textbf{\textit{Description of User}}& \textbf{\textit{Mentions}} \\
			\hline
			@NARENDRAMODI & Mr. Narendra Modi, PM of India & \textit{14,160}\\ 
			\hline
			@ARUNJAITLEY & Mr. Arun Jaitley, FM of India & \textit{8,675} \\
			\hline
			@PMOINDIA & Office of PM &\textit{3,661}\\
			\hline
			@PIYUSHGOYAL & Mr. Piyush Goyal, Union Minister & \textit{3,641} \\
			\hline
			@INCINDIA & Indian National Congress &\textit{3,183}\\
			\hline
			@FINMININDIA &Ministry of Finance &\textit{3,176} \\
			\hline
			@OFFICEOFRG &Rahul Gandhi, President, INC &\textit{3,086}\\
			\hline
			@ASKGST\_GOI  &GOI official Handle for GST Queries&\textit{3,004} \\
			\hline
			@BJP4INDIA &Bharatiya Janata Party &\textit{2,916}\\
			\hline
			@ADHIA03 &Dr Hasmukh Adhia, Fin. Sec., GOI &\textit{2,815} \\
			\hline
			@SAPINDIA &SAP India Pvt. Ltd. &\textit{2,194}\\
			\hline
			@ARVINDKEJRIWAL & Mr. Arvind Kejriwal, CM, Delhi &\textit{2,064} \\
			\hline
			@SANJAYAZADSLN &Mr. Sanjay Singh, Rajya Sabha MP, AAP &\textit{1,873}\\
			\hline
			@GST\_COUNCIL &GST Council &\textit{1,808} \\
			\hline
			@AMITSHAH &Mr. AMIT Shah, President of BJP &\textit{1,656}\\
			\hline
			@PREETISMENON &Ms. Preeti Menon, Member, AAP &\textit{1,652} \\
			\hline
			@AJAYMAKEN &Mr. Ajay Maken, President Delhi Congress &\textit{1,628}\\
			\hline
			@MSISODIA &Mr. Manish Sisodia, Deputy CM of Delhi &\textit{1,601} \\
			\hline
			\hline
			\multicolumn{2}{c}{Total Mentions} & \multicolumn{1}{c}{\textit{62,793}} \\
			\hline
			\hline
		\end{tabular}
		\label{tab1}
	\end{center}
\end{table}
\begin{figure*}[!t]
	\subfloat[Hashtag Word Cloud]
	{\includegraphics[width=6.1 cm,height=5.9cm]{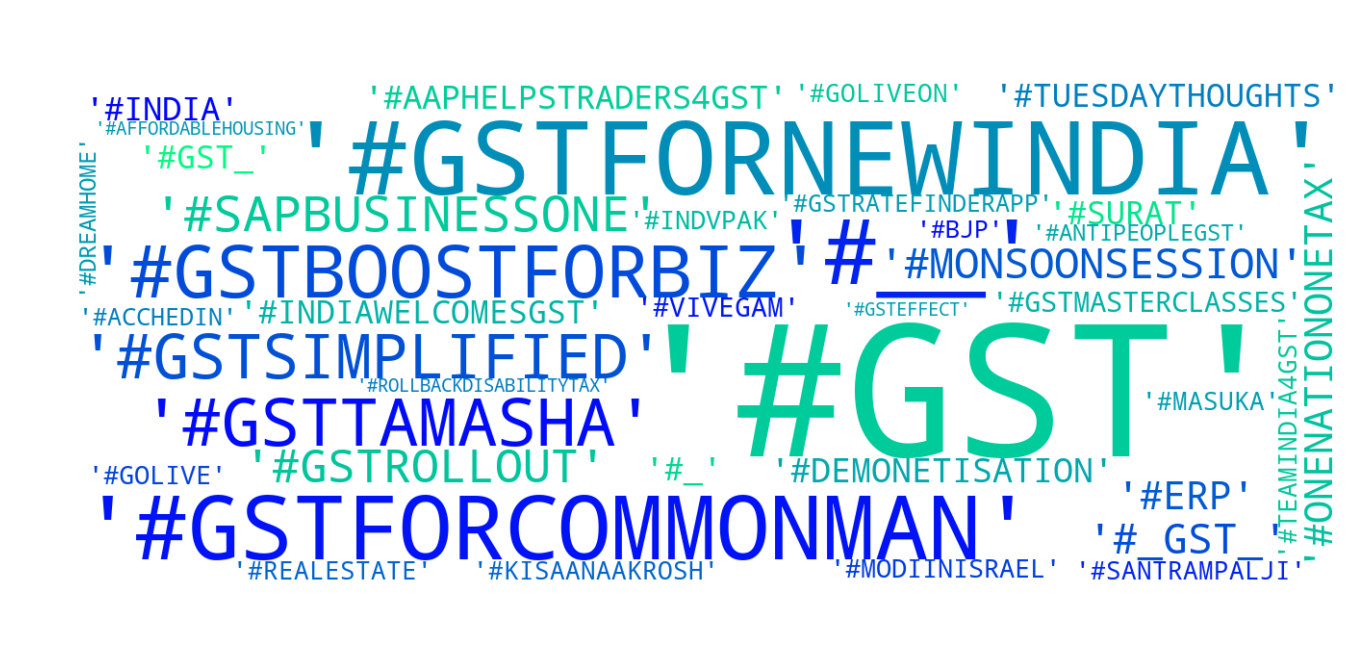}%
		\label{f1}}
	\subfloat[Mention @user Word Cloud]
	{\includegraphics[width=6.1 cm,height=5.9cm]{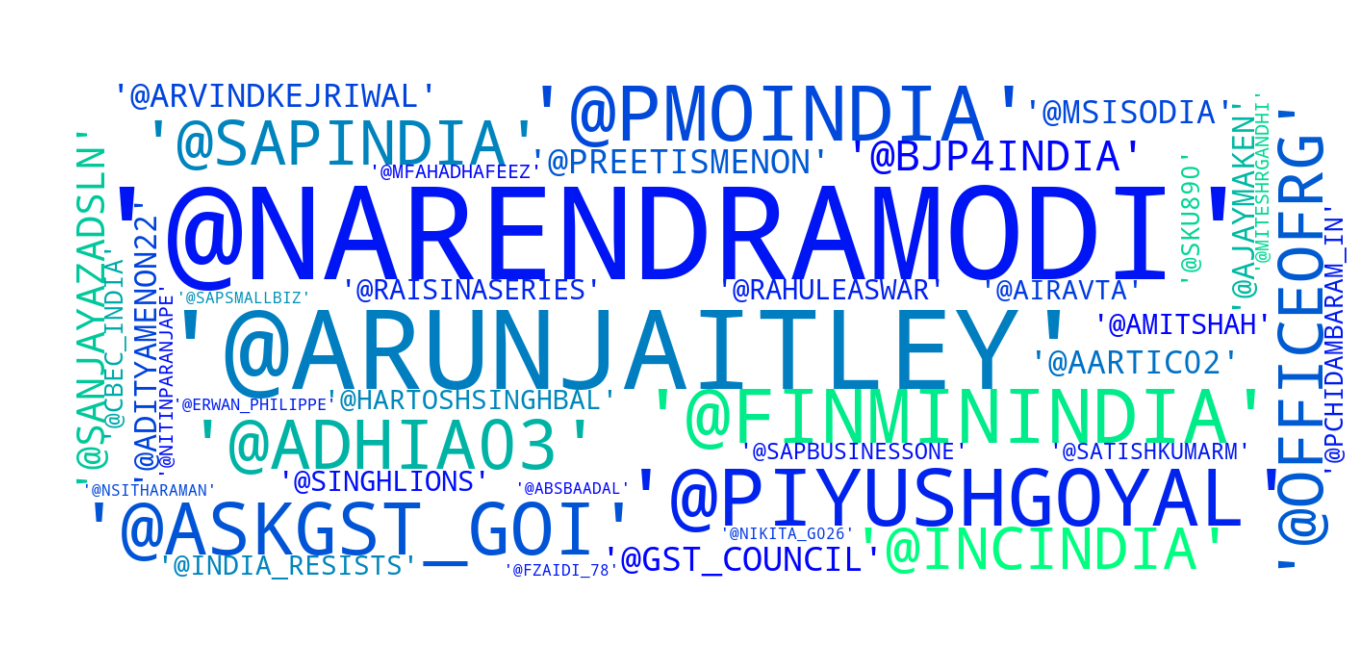}%
		\label{f2}}
	\caption{Showing the Word Cloud for Top 40 Hashtags and Top 40 Mentions}
\end{figure*}
\begin{figure*}[!t]
	\centering
	\subfloat[Sentiments on Mr. Modi Tweets]
	{\includegraphics[width=6.3 cm,height=4cm]{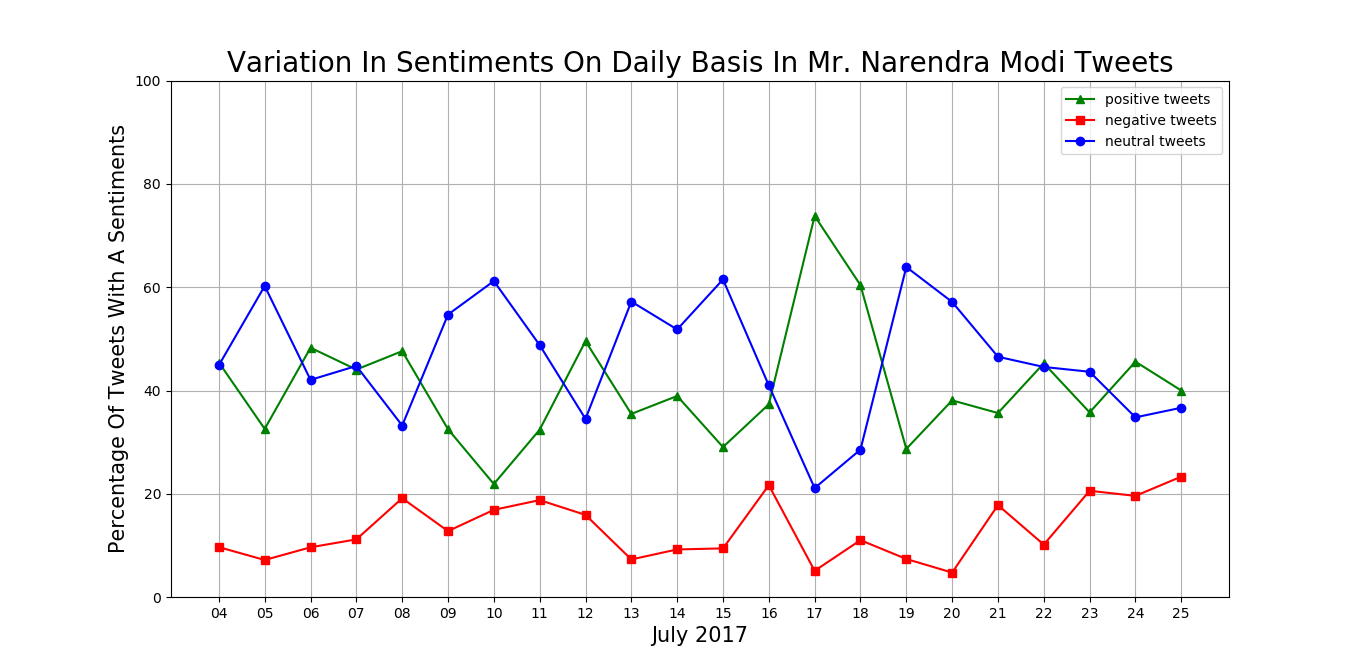}%
		\label{f1}}
	\subfloat[Emotions on Mr. Modi Tweets]
	{\includegraphics[width=6.3 cm,height=4cm]{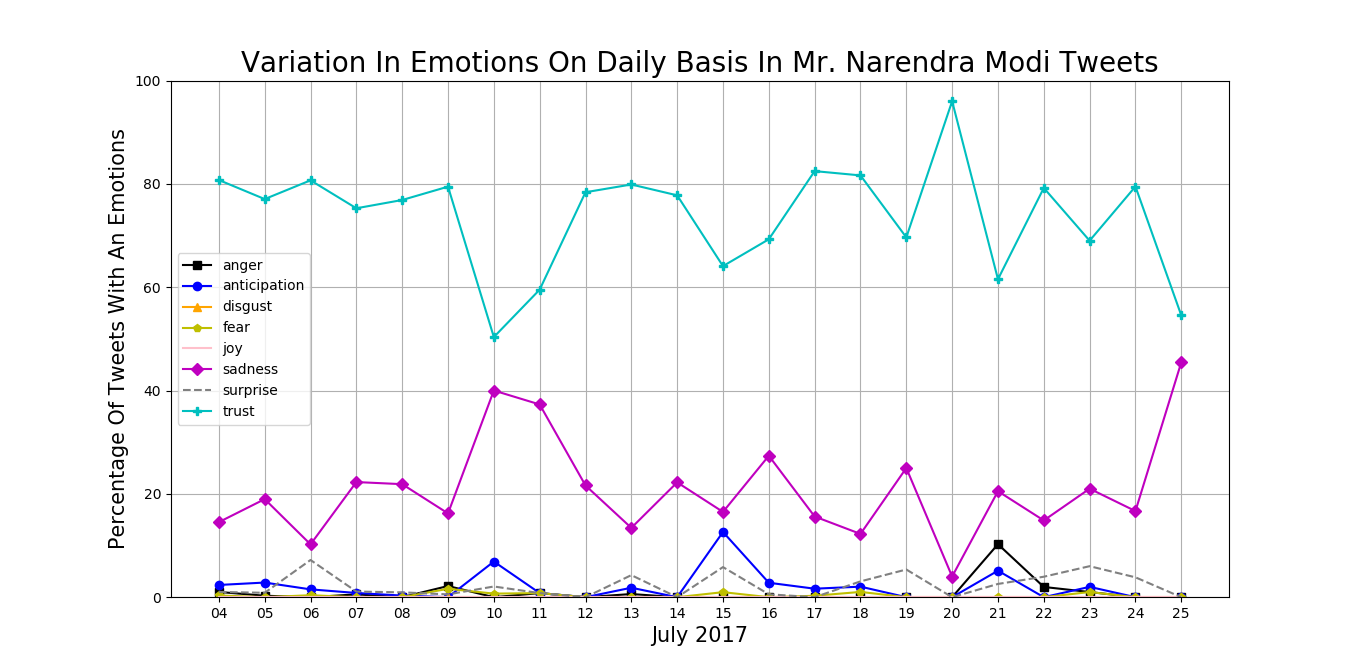}%
		\label{f2}}
	\\
	\subfloat[Sentiments on Mr. Arun Jaitley Tweets]
	{\includegraphics[width=6.3 cm,height=4cm]{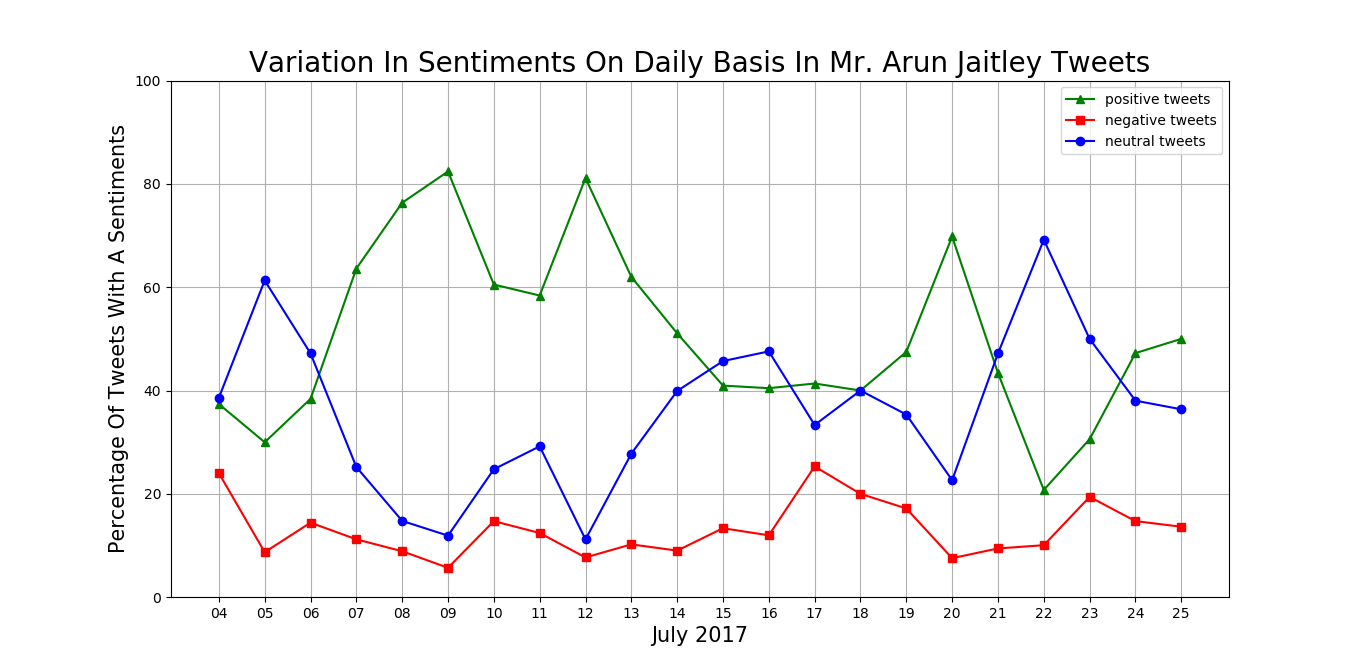}%
		\label{f2}}
	\subfloat[Emotions on Mr. Arun Jaitley Tweets]
	{\includegraphics[width=6.3 cm,height=4cm]{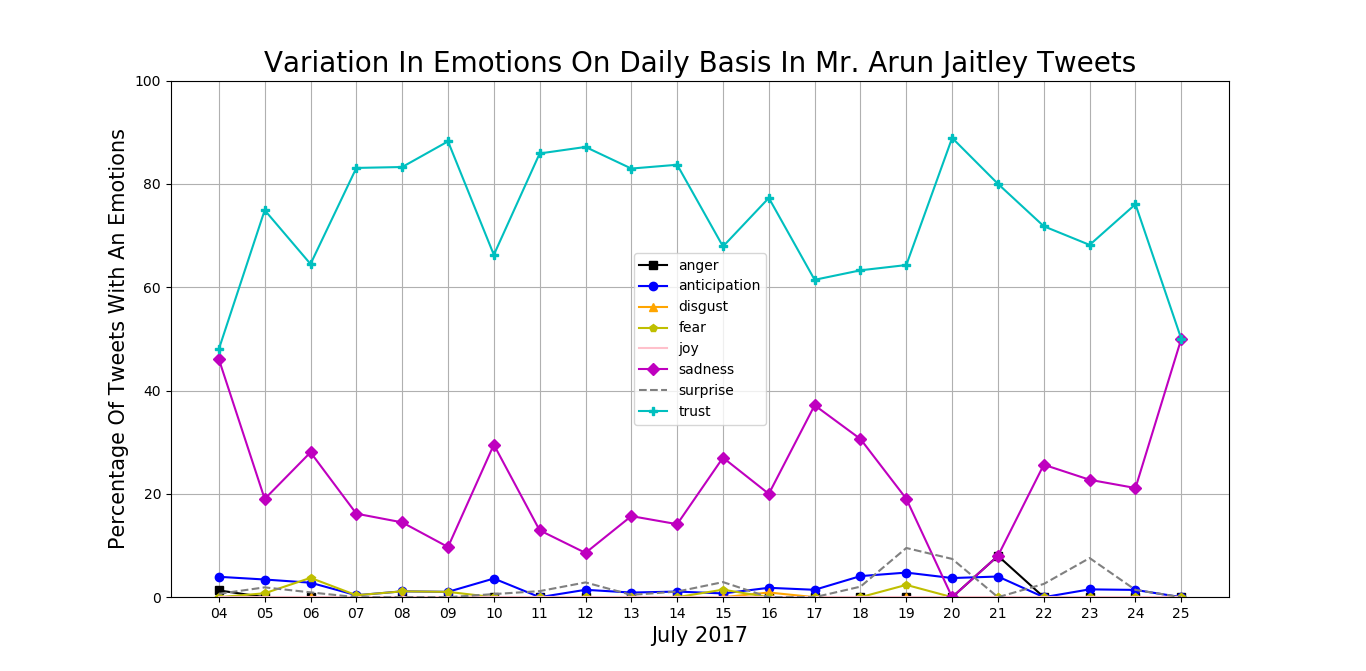}%
		\label{f2}}
	\caption{Showing the Variation of Sentiments and Emotions On Tweets Addressed To Mr. Narendra Modi and Mr. Arun Jaitley}
\end{figure*}
\begin{table}[tb]
	\caption{Location Wise Number of Tweets}
	\begin{center}
		\begin{tabular}{|c|c|c|c|}
			\hline
			\cline{2-4} 
			\textbf{Location} & \textbf{\textit{No of Tweets}}& \textbf{\textit{Location}}& \textbf{\textit{No of Tweets}} \\
			\hline
			Delhi-NCR & \textit{10919}& Maharasthra & \textit{8437} \\
			\hline
			Karnataka & \textit{3583}& Gujrat & \textit{3146} \\
			\hline
			Tamilnadu & \textit{3067}&  Rajasthan & \textit{2548} \\
			\hline
			Uttar Pradesh & \textit{2586}& Madhya Pradesh & \textit{2552} \\
			\hline
			Haryana & \textit{1047}& Punjab & \textit{1225} \\
			\hline
			West Bengal & \textit{1204}& Chhattishgarh & \textit{649} \\
			\hline
			Bihar & \textit{590}& Jammu \& Kashmir & \textit{520} \\
			\hline
			Uttarakhand & \textit{361}& Orissa & \textit{371} \\
			\hline
			Jharkhand & \textit{362}& Kerala & \textit{303} \\
			\hline
			Goa & \textit{167}& Assam & \textit{165} \\
			\hline
			Telangana & \textit{95}& Andhra-Pradesh & \textit{243} \\
			\hline
			Himachal Pradesh & \textit{108} &  \\
			\hline
			\multicolumn{2}{c}{India (State Not Mentioned )} & \multicolumn{2}{c}{\textit{11,524}} \\
			\hline
			\hline
			\multicolumn{2}{c}{India (Total Tweets)} & \multicolumn{2}{c}{\textit{55,773}} \\
			\hline
			\hline
			\multicolumn{2}{c}{Foreign} & \multicolumn{2}{c}{\textit{2840}} \\
			\hline
			\hline
			\multicolumn{2}{c}{Total Tweets} & \multicolumn{2}{c}{\textit{58,613}} \\
			\hline
			\hline
		\end{tabular}
		\label{tab1}
	\end{center}
\end{table}
We also analysed the tweets on an hourly basis for which we divided every-day's time into 24 slots. Each slot is of 1 hour, where the 1st slot starts at 00:00:00 Hrs and ends at 00:59:59 Hrs. Accordingly, the 24th slot starts at 23:00:00 Hrs and ends at 23:59:59 Hrs.
We summed up the count of tweets in each slot over 22 days of period to find out the peak time of the day when maximum people were tweeting. It can be seen in the graph in Fig-2(c), that most of the tweets have been posted from 07:00:00 Hrs to 19:00:00 Hrs and peak time slot is from 17:00:00 Hrs to 18:00:00 Hrs.

We have plotted word cloud for hashtags and most of the hashtags indicate that the taxation system has been simplified and will boost the Indian economy. Hashtag word cloud in Fig-3(a) shows that the top hashtags in the GST data are \#GST, \#GSTFORCOMMONMAN, \#GSTBOOSTFORBIZ, \#GSTFORNEWINDIA, \#GSTSIMPLIFIED, \#ONENATIONONETAX, \#INDIAWELCOMESGST which clearly indicates that public is expressing trust and positive sentiments towards modified tax policy of India.

We have also extracted the mentions @user{Table 2}, from the tweets and calculated their counts in the entire dataset. This analysis has been performed to find out the personalities that are being addressed by more number of people. We have also plotted word cloud for the top forty mentions in Fig-3(b) so that it's easy to visualize and find out the top mentions during the event.
From this analysis we have found out that more than 15,400 users have been mentioned 2,77,200 times in totality in the collected GST twitter dataset. Prime Minister of India, Mr. Narendra Modi has been mentioned 14,160, highest number of times, followed by Finance Minister of India, Mr. Arun Jaitley.
We have not shown the data of 15,390 users as their count of mentions is less than users specified in {Table 2}. 

We have plotted the sentiment and emotion analysis graphs for the top two leaders, Mr. Narendra Modi and Mr. Arun Jaitley. All the graphs show that tweets mentioning these two personalities are highly positive in sentiment and highly trustworthy in emotion.

\subsubsection{Spatial Analysis}
In Spatial analysis, we have mined the data to extract the locations, as per their availability, from the tweets and analysed the count and sentiments of tweets for the cities with maximum number of tweets.

We have also extracted the tweet count and the sentiments which are posted from abroad. This shows that people outside India also took interest in this process and presented their views on the new tax regime. We are assuming that most of the tweets has been posted by the NRI's who are concern about their country.

Table 3 shows the state wise tweets count, which shows that the maximum number of tweets has been posted from the Delhi-NCR region, national capital of India.
We have extracted approximately 80000 tweets, that are having some record of the locations but finally we extracted approximately 59,000 tweets mentioning a real location.

Those who have not mentioned the name of any state or a city but are from India, have been shown under a separate tag - \textbf{INDIA}.

Table-3 shows that most of the tweets were posted from the metropolitan cities. It could have been due to following reasons - other people are not interested in the event or they are not expressing their views on social media. People from other cities/towns don't have much access to the social media or Internet, or they don't have much information about the event. That's why they are unable to give their take on the issue.
Out of the 58,613 tweets more than 70\% of the tweets are coming from cities like Delhi-NCR, Mumbai, Bangalore, Chennai, Gujrat, Hyderabad, Pune which shows that social media is dominated by small group of people and hence the views and opinions on social media are generated by a small section of the society.
The major challenges in analysing twitter or any other social media platform is get the true sentiment, to know whether these groups or sections of people are representing the views of the whole country or not.

\section{CONCLUDING REMARKS AND FUTURE RESEARCH}
In this paper we analysed the GST twitter data on the basis of time and location. We performed temporal analysis to show the varying amount of sentiment and emotions of public after GST implementation in India. The analysis was performed on the POST-GST data which we started collecting three days after its implementation and continued it for next twenty-two days. We have found out that most of the tweets represent positivity and trust. We have also found out that most of the people are addressing the PM of India, Mr. Narendra Modi and the Finance Minister of India, Mr. Arun Jaitley in their tweets expressing positive sentiments and trust in their emotions.
We have also performed temporal analysis to find out the peak hours of the day when most of the people are expressing their views on twitter and it came out to be 17:00:00 Hrs to 18:00:00 Hrs IST.

In spatial or location based analysis it has been observed that more than 45\% of the tweets have been tweeted from Delhi-NCR \& Mumbai. The top six states, as can be seen from the table-3, accounts for approximately 70\% tweets.

The whole procedure shows that the data on which analysis has been performed comes from a small part of the country which is comparatively more developed. It might have due to many reasons like other people are not interested in expressing their views, don't have enough knowledge about the event, or don't have Internet connectively.

In this research we have assumed that all the users are genuine. To refine this research further, we shall try to find out and remove the redundancy and fake user data in the future works we do.

\bibliographystyle{JISEbib}
\bibliography{references}

\end{document}